\documentclass[12pt]{article}

\usepackage{scicite}

\usepackage{times}
\usepackage{physics} 
\usepackage{braket}
\usepackage{xcolor}

\newcommand{\black}[1]{\textcolor{black}{#1}}

\newenvironment{sciabstract}{%
\begin{quote} \bf}
{\end{quote}}

\usepackage{graphicx}
\usepackage{lscape}
\usepackage{multirow}

\newcounter{lastnote}

\title{Optically-Sampled Superconducting-Nanostrip Photon-Number Resolving Detector for Non-Classical Quantum State Generation}

\author
{Mamoru Endo,$^{1,2,\ast}$ Kazuma Takahashi,$^{1}$ Takefumi Nomura,$^{1}$ Tatsuki Sonoyama,$^{1}$  \\
Masahiro Yabuno,$^{3}$ Shigehito Miki,$^{3}$ Hirotaka Terai,$^{3}$ \\
Takahiro Kashiwazaki,$^{4}$ Asuka Inoue,$^{4}$ Takeshi Umeki,$^{4}$ \\
Rajveer Nehra,$^{1,5,6,7}$ Kan Takase,$^{1,2}$ Warit Asavanant,$^{1,2}$ Akira Furusawa$^{1,2,**}$\\
\\
\normalsize{$^{1}$Department of Applied Physics, School of Engineering, The University of Tokyo,}\\
\normalsize{7-3-1 Hongo, Bunkyo, Tokyo 113-8656, Japan}\\
\normalsize{$^{2}$Optical Quantum Computing Research Team, RIKEN Center for Quantum Computing,}\\
\normalsize{2-1 Hirosawa, Wako, Saitama 351-0198, Japan}\\
\normalsize{$^{3}$Advanced ICT Research Institute, National Institute of Information and Communications Technology,}\\
\normalsize{588-2 Iwaoka, Nishi, Kobe 651-2492, Japan}\\
\normalsize{$^{4}$NTT Device Technology Labs, NTT Corporation,}\\
\normalsize{3-1 Morinosato Wakamiya, Atsugi, Kanagawa 243-0198, Japan}\\
\normalsize{$^{5}$Department of Electrical and Computer Engineering, University of Massachusetts Amherst,}\\
\normalsize{Amherst, Massachusetts 01003, USA}\\
\normalsize{$^{6}$Department of Physics, University of Massachusetts Amherst,}\\
\normalsize{Amherst, Massachusetts 01003, USA}\\
\normalsize{$^{7}$College of Information and Computer Science, University of Massachusetts Amherst,}\\
\normalsize{Amherst, Massachusetts 01003, USA}\\
\\
\normalsize{To whom correspondence should be addressed; *E-mail:  endo@ap.t.u-tokyo.ac.jp }\\
\normalsize{**E-mail:  akiraf@ap.t.u-tokyo.ac.jp}
}

\date{}

\begin{document} 
\baselineskip24pt
\maketitle 
\begin{sciabstract}
Photon number-resolving detectors (PNRDs) are the ultimate optical sensors. Superconducting-nanostrip photon detectors (SNSPDs), traditionally known as ON-OFF detectors, have recently been found to have photon number resolving capability without multiplexing. This discovery positions them to become true PNRDs. However, their practical use is limited by the need to precisely detect tiny signal differences with low signal-to-noise ratios within sub-nanosecond time frames. We overcome this challenge using optical sampling with a dual-output Mach Zehnder modulator (DO-MZM) and ultra-short pulsed laser. By adjusting the DO-MZM's bias voltage to nearly balance the outputs, this method enables sensitive detection of picosecond-order signal differences, achieving a temporal resolution of 1.9 ps and facilitating real-time photon number resolution. We applied this method to produce various non-classical quantum states, enhancing their non-classicality through photon number resolution. This advancement marks a significant shift from principle verification to practical application for SNSPD-type PNRDs in diverse quantum optics fields.
\end{sciabstract}


\section*{Introduction}
In quantum optics, photon number-resolving detectors (PNRDs) emerge as one of the ultimate sensors \cite{Lita2022}. 
Various types of ON-OFF detectors (or single-photon detectors), such as avalanche photodiodes \cite{Cova1996}, and superconducting nanostrip photon detectors (SNSPDs) \cite{Reddy2020}, have been researched and commercialized. 
Pseudo PNRDs, constructed by temporally \cite{Jonsson2020} or spatially \cite{Divochiy2008,Cheng2022} multiplexing these ON-OFF detectors are suitable for applications like weak light power sensors, where absolute photon number information is not required. 
In contrast, particularly in the rapidly advancing field of optical quantum information processing, measuring the absolute photon number is essential and pseudo PNRDs are not suitable, unfortunately \cite{Jonsson2019,Provaznik2020}. 
To measure absolute photon number by pseudo PNRDs, it is necessary to parallelize a large number of ON-OFF detectors with nearly 100\% detection efficiency, ensuring that no more than one photon hits each detector simultaneously \cite{Nehra2020}. However, this approach is not practical. \black{For instance, even with eight ON-OFF detector array, each detector having 100\% efficiency and no dark counts, the maximum resolvable photon number is limited to three. When the detection efficiency is reduced to a more realistic value of 95\%, arranging as many as 32 ON-OFF detectors only allows for the discrimination of at most five photons \cite{Jonsson2019}.} Also, it faces significant technical challenges, including wiring complexities, the limited cooling capacity of refrigerators, and reduced temporal resolution \cite{Kruse2017}.
PNRDs with true photon number resolving capability without any multiplexing would enable logical qubit generation \cite{Gottesman2001} and nonlinear quantum operations \cite{Sakaguchi2023}, which would greatly advance the realization of fault-tolerant optical quantum computers with high-clock frequency \cite{Takeda2019}.

Superconducting transition edge sensors (TES) stand out as true PNRDs \cite{Lita2022, Fukuda2019}. They measure photon energy or count based on the temperature rise when photons are absorbed by a superconducting thin film such as titanium, tungsten, iridium and so on. TES offers exceptional sensing capabilities with high detection efficiency across a broad wavelength range, low dark counts, and the ability to discern over 20 photons. Indeed, TES has enabled the generation of non-classical quantum states \cite{Gerrits2010,Endo2023} with negative Wigner functions. However, TES falls short in response speed and temporal resolution due to electronic circuit constraints in signal readout. Also, their low operational temperature around 100 mK necessitates complex refrigeration systems, presenting a high barrier for practical use.

Recently, SNSPDs have been suggested to have photon number-resolving capabilities even without multiplexing \cite{Cahall2017,Zhu2020,Kong2024}. Utilizing the distinct rise in the signal corresponding to the photon detection, these detectors show high potential for practical application without needing a special structure. Detector tomography has confirmed their capability to discern a few photons \cite{Endo2021}, with several applications already demonstrated \cite{Sempere-Llagostera2022}. The temporal resolution of SNSPD is well below 100 ps and is promising as a fast PNRD \cite{Korzh2020}. \black{However, SNSPD-type PNRD requires instruments with a temporal resolution of better than several tens of picoseconds, making real-time photon number resolving particularly challenging.}

In this paper, we propose a solution to this challenge using optical sampling with a pulsed laser and a dual-output Mach-Zehnder modulator (DO-MZM). By adjusting the bias voltage of the DO-MZM to nearly balance the output, this method enables the sensitive detection of slight differences in the signal, facilitating real-time photon number resolution. \black{This method easily facilitates real-time 1.9-ps temporal resolution measurements, making fast and real-time photon number-resolving with an SNSPD feasible. }
We applied this technique to generating non-classical quantum states, which is crucial for optical quantum information processing. We generated squeezed light by exciting a waveguide optical parametric amplifier (WG OPA), part of which we measured with the SNSPD-type PNRD, using a photon subtraction method to produce non-classical states \cite{Danka1997}. These states, approximating Schr\"{o}dinger cat states, serve as essential resources for recently realized Gottesman-Kitaev-Preskill (GKP) qubits in optical systems \cite{Gottesman2001, Konno2024}, which is a strong candidate for logical qubits. After measuring the quadrature-phase amplitudes with a homodyne detector (HD), we performed quantum state tomography, obtaining density matrices and Wigner functions for the evaluation. 
Thanks to the high temporal resolution of the SNSPD-type PNRD, we could eliminate the influence of unnecessary detection events, improving the generated states. Furthermore, we managed to generate a quantum state, that is two-photon subtracted state, unattainable without PNRDs.

The SNSPD-type PNRD and optical sampling can easily integrate into existing experimental apparatus, offering various applications beyond this study. It promises to accelerate the development not only in optical quantum information processing but also in broader optical research.

\section*{Results}
\subsection*{Optical sampling method}
\begin{figure}[ht]
    \centering
    \includegraphics[width=1\linewidth]{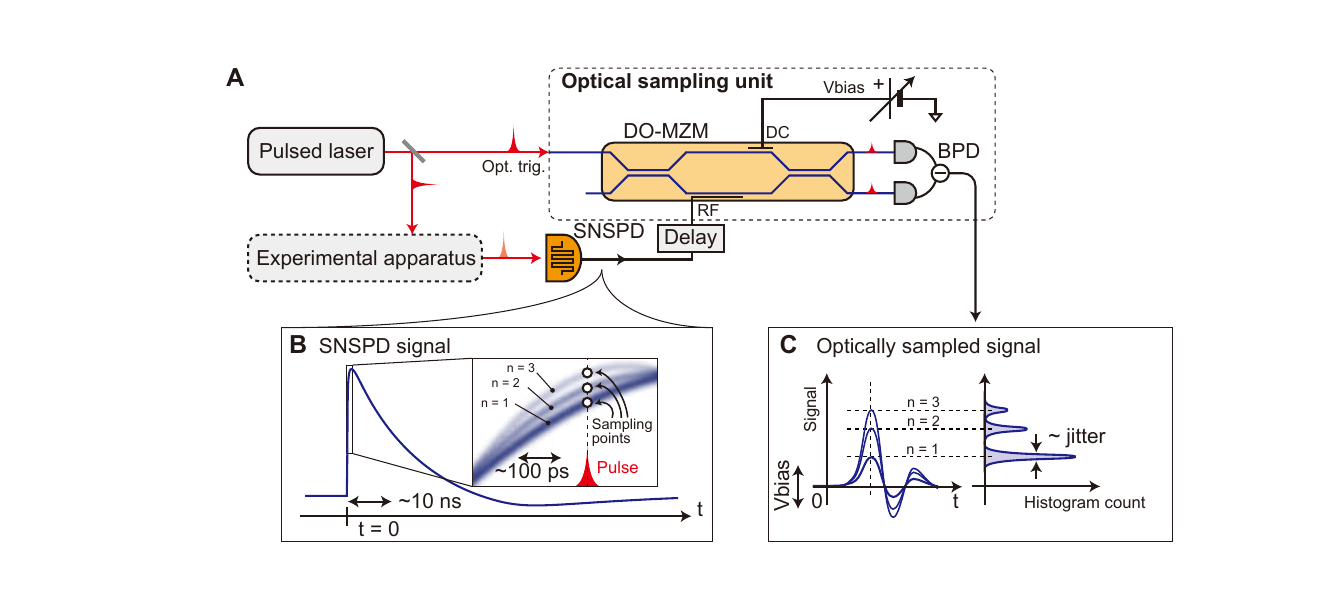}
    \caption{Concept diagram of the SNSPD-type PNRD and the optical sampling method. A. An example of an experimental setup using an SNSPD-type PNRD with optical sampling. The pulse laser is split into two, with one part used for quantum optics experiments and the resultant ultra-weak light is detected by the SNSPD. The electrical signal from the SNSPD, after appropriate amplification, is inputted into the RF port of a dual-output Mach-Zehnder modulator (DO-MZM). An RF delay line is used for adjusting the sampling point. The other part of the pulse laser is fed into the optical input port of the DO-MZM. As mentioned in the main text, the DC bias of the DO-MZM is adjusted as necessary. B. An image of the signal obtained from the SNSPD. Typically, the signal has a rise time of less than a nanosecond and a fall time of several tens of nanoseconds. An expanded view of the rising part is shown in the inset. The signal separates in a region of about a hundred picoseconds, corresponding to photon numbers 1, 2, and 3, respectively. By appropriately adjusting the timing of the light pulse in the setup of A, a signal example obtained by the optical sampling method is shown in C. It provides the voltage value of the SNSPD's signal at the time of optical sampling\black{, from which one can know the number of photons}. To the right is a histogram of the voltage at a certain time. The width of the histogram is influenced by electrical noise, light noise, etc., but is largely due to the jitter of the SNSPD.
    }
    \label{fig:concept}
\end{figure}

In our SNSPD-type PNRD, photon number resolution is performed based on slight differences in the signal's rising edge \cite{Cahall2017,Endo2021}. The experimental setup for this purpose is shown in Fig.~\ref{fig:concept} A. The output of a pulsed laser is split into two, with one part directed to the optical input port of a DO-MZM and the other to a setup for quantum optics experiments (like quantum state generation as in this paper, weak light LIDAR, fluorescence measurement, etc.). The target ultra-weak light is detected by the SNSPD installed inside a refrigerator. Since the signal from the SNSPD is generally weak, it is usually amplified using low-noise cryogenic amplifiers in the refrigerator. The amplified signal is then fed into the RF input port of the DO-MZM via an RF delay circuit. The signal from the SNSPD typically appears as shown in Fig.~\ref{fig:concept} B. Depending on the structure of the SNSPD and readout circuits, the rise time is usually less than one nanosecond, with a fall time of several tens of nanoseconds. An expanded view of the signal over a few hundred picoseconds is shown in the inset, illustrating cases where the photon number is 1, 2, and 3. In this manner, the signal separates in the region of about a hundred picoseconds according to the photon number, necessitating high-temporal-resolution instruments in general. Our proposed optical sampling method uses an RF delay line to adjust the timing of the light pulse as shown in Fig.~\ref{fig:concept} B, allowing us to optically sample the SNSPD signal with high temporal resolution. A low-noise voltage source applies a DC bias to the bias port of the DO-MZM, adjusting the balance of the DO-MZM's optical output. The output of the DO-MZM is detected by a balanced photo-detector (BPD). An example of a signal obtained by the optical sampling method is shown in Fig.~\ref{fig:concept} C. Signals with distinctly different peak values according to the photon number are obtained. A histogram of the voltage values at a certain time is shown on the right.

Since the width of the histogram of the signal obtained by the optical sampling method mainly contains information on the timing jitter of the SNSPD, it is also possible to measure the timing jitter of the SNSPD if the measurement limit of the optical sampling method is sufficiently small. When measuring timing jitter, the sensitivity of the measurement can be increased by setting the bias voltage so that the signal from one-photon detection is near zero and increasing the optical power and BPD's transimpedance gain.
To improve the measurement's sensitivity and temporal resolution, we can increase the power of the pulse laser and optimize the transimpedance gain of the BPD. Using a conventional (single-output) MZM and a single photodetector would easily saturate the photodetector's output. Here, we use a DO-MZM and BPD, which allows us to operate under conditions where the two optical outputs are nearly balanced by tuning the bias voltage, making it easier to improve the sensitivity and temporal resolution. By preparing one DO-MZM and one BPD, it is easily possible to upgrade from the conventional method to the proposed optical sampling method.

\begin{figure}[ht]
    \centering
    \includegraphics[width=1\linewidth]{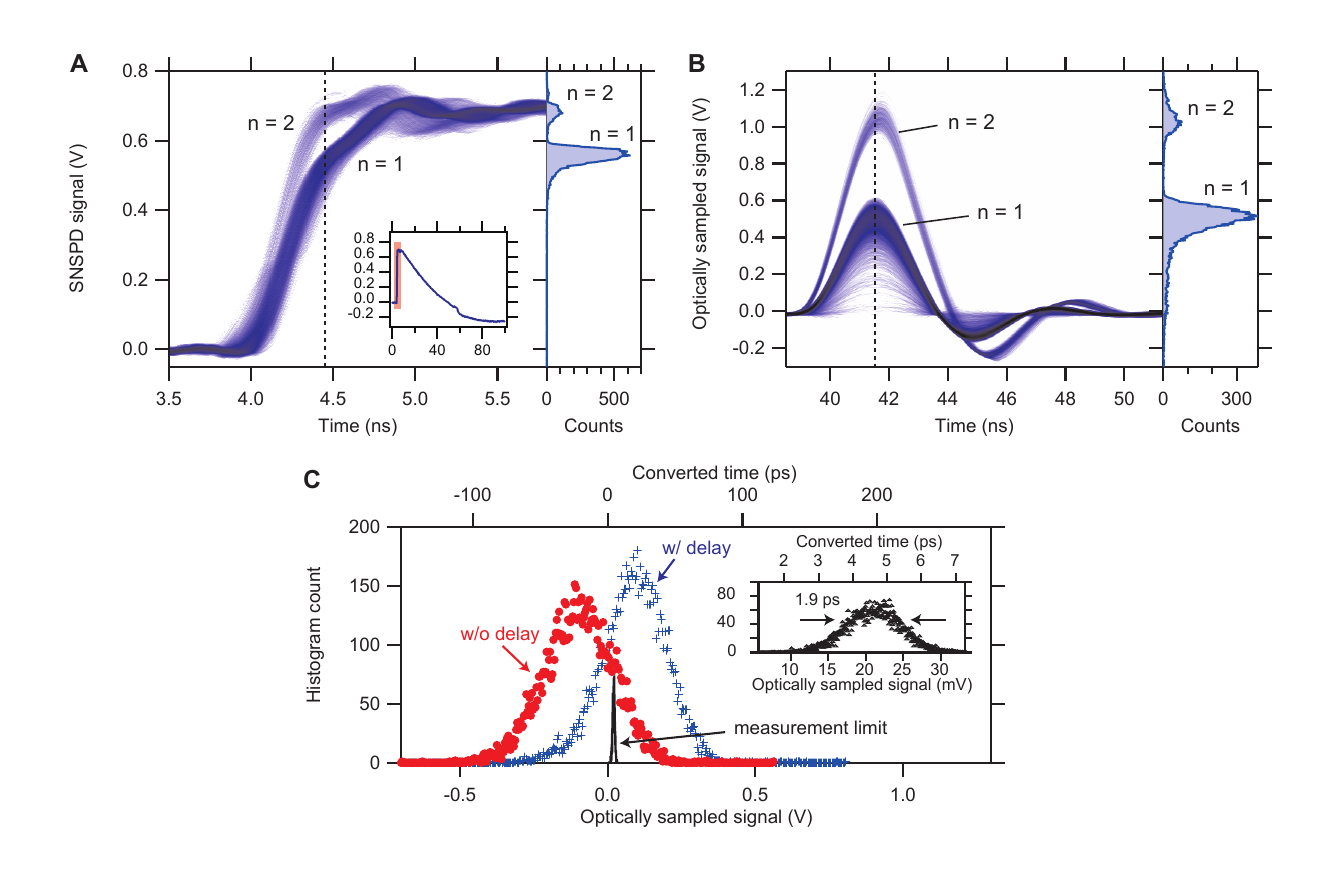}
    \caption{Actual signals from the SNSPD. A. The conventional method. The inset shows the whole signal from the SNSPD and the main plot shows the extended plot. The right plot shows the voltage histogram at the dashed line. B. Optical sampling method. The shows the voltage histogram at the dashed line. C. Timing jitter and measurement limitation estimations of the optical sampling method. Red, blue, and black traces show the voltage histogram data without the delay, with the delay, and the measurement limit, respectively. Note that the optical power, amplifier's gain, and bias voltage were different when the data in A and B were acquired and when the data in C was acquired.}
    \label{fig:snspd_signal}
\end{figure}

Figure~\ref{fig:snspd_signal} A and B show the experimental comparison between the conventional and the optical sampling methods. We split a laser with a repetition frequency of 5 MHz and a pulse duration of about 2 ps, using one of the splits as ultra-weak coherent light with an average photon number of less than one per pulse and injected it into an SNSPD (see Methods). The SNSPD was cooled to a temperature of 700 mK and biased with a current of about 24 $\mathrm{\mu A}$. The signal was amplified by a low-noise, cryogenic amplifier (Cosmic Microwave Technologies, Inc., CITLF3-20K) cooled to 3.2 K. A superconducting NbTi coaxial cable (COAX CO., Ltd.) is used between the SNSPD and the cryogenic amplifier to avoid heat inflow. Then the \black{signal} is further amplified by a room-temperature amplifier (Pasternak, PE15A10007). Figure~\ref{fig:snspd_signal} A plots the signal from the SNSPD captured by an oscilloscope (Keysight, DSOS204A), corresponds to Fig.~\ref{fig:concept} B. The inset shows the entire signal from the SNSPD, with the red-framed part magnified in the main figure. Here, we have overlaid about 16,000 waveforms. We plot a voltage histogram on the right at a certain time (dashed line). As can be seen from both plots, the signal is discretized, allowing for photon number-resolving of one or two photons. However, as the rise time of this signal is less than 500 ps and the difference is small, measurement devices with high temporal resolution are essential. In the SNSPD used this time, components of three or more photons were not observed, but with appropriate devices, it is possible to detect more photons.

We simultaneously measured the same signal with the optical sampling method as shown in Fig.~\ref{fig:concept} A. We took a part of the pulsed laser, set its average power to about 1 mW, and injected it into a DO-MZM (EOSpace, AX-2x2-0MSS-10-PFA-PFA-LV). The two outputs were received by a balanced photodetector (Hamamatsu, C12688-2). The output signal was captured by the oscilloscope. 
Figure~\ref{fig:snspd_signal} B shows both the waveforms acquired by the oscilloscope and the voltage histogram at a certain time (dashed line). It is clear that photon number resolving is distinctly possible through the optical sampling method. An important aspect here is that, in optical sampling, the information about the photon number directly manifests as differences in the signal's peak values, which is particularly advantageous for applications requiring real-time photon number information. For instance, by setting the oscilloscope's trigger voltage to 0.7 V, events involving two photons can be easily extracted.

We also investigated the temporal resolution of the optical sampling method. To measure a conversion factor from the voltage to time, a known delay line was inserted between the SNSPD and the DO-MZM. The signal output by the optical sampling method varied with the magnitude of the delay. From the voltage change and delay, the conversion factor (in the unit of V/s) can be calculated. In our case, we used a variable RF phase shifter (Sage Labs, 6705-2) as the delay line, applying a delay of 41.7 ps. Histograms of the output of the optical sampling method without and with the delay (red: without delay, blue: with delay) are shown in Fig.~\ref{fig:snspd_signal} C. The central voltage change in the histogram was 201 mV. Thus, the conversion factor was calculated to be $4.81\times10^{9}$ V/s, and the converted time is shown on the top axis of the figure. The histogram's width corresponds to the jitter of the SNSPD, which was $60.3(5)$ ps (full-width at half maximum) in this experiment. This value is consistent with the jitter measured by the oscilloscope. 
Note that the red and blue histograms have different widths because the slope varies with the sampled time. The estimated jitter value shown here is the worst case.  Next, we measured the signal of the BPD when the SNSPD's signal was blocked (black markers in Fig.~\ref{fig:snspd_signal} C). Dividing the histogram's width by the gain provided the resolution of the optical sampling method, which was found to be \black{$1.90(3)$} ps, which is comparable to the highest performance time-correlated single photon counting modules \cite{Braun2023} \black{and sufficiently small to resolve the photon number}. As discussed in the text, this value is limited by the pulse width, power of the laser used, and the transimpedance gain and noise of the BPD.

\subsection*{Nonclassical quantum state generation}
We next applied the SNSPD-based PNRD to the generation of non-classical quantum states by photon subtraction, where real-time photon-number resolving measurement is essential. 
This section explains the concept of photon subtraction followed by experimental details. 
\subsubsection*{Photon subtraction and state verification}
Photon subtraction was conceived for generating approximate superpositions of coherent states: $\ket{\alpha}+e^{i\theta}\ket{-\alpha}$, or Schr\"{o}dinger cat states \cite{Danka1997}. It involves picking off a small portion of the output of a squeezed vacuum $\hat{S}(r)\ket{0}$, generated by an optical parametric amplifier (OPA), via a beamsplitter and measuring it with a PNRD. Here, $\hat{S}(r)=\exp\left[\frac{r}{2}(\hat{a}^2-(\hat{a}^\dagger)^2)\right]$ is a squeezing operator, and $r$ is a squeezing parameter, $\hat{a}^\dagger,\hat{a}$ are creation and annihiration operators, respectively. Depending on the PNRD's output $n$, a state $\hat{a}^n\hat{S}(r)\ket{0}$ is produced at the other output of the beamsplitter. This process appears as if $n$ photons are being subtracted from the squeezed vacuum, hence the name. It is important to note that the squeezed vacuum is a superposition of even-numbered photon states, so the state's parity produced after subtracting $n$ photons matches the parity of $n$. The sign of the value at the origin of the Wigner function, calculated from the density matrix, corresponds to the parity \cite{Royer1977}; therefore, when $n$ is odd, the value at the origin is negative, and when $n$ is even, it is positive. While numerous experiments have been conducted using photon subtraction, most are limited to the case of $n=1$ due to the difficulty in photon number-resolving measurements \cite{Ourjoumtsev2006,Neergaard-Nielsen2008,Asavanant2017,Baune2017,Ra2019,Takase2022}. Only a few experiments have detected multiple photons, using Transition Edge Sensors (TES) \cite{Gerrits2010,Endo2023} or single-photon detector arrays \cite{Ourjoumtsev2006}.

State verification is performed through homodyne measurement, which involves interfering the target light with local oscillator light (LO) at a beamsplitter at phase $\theta$ and detecting the outputs with a balanced photodiode. The output current corresponds to the quadrature phase amplitude $\hat{x}_\theta=\frac{\hat{a}e^{-i\theta}+\hat{a}^\dagger e^{i\theta}}{\sqrt{2}}=\hat{x}\cos\theta + \hat{p}\sin\theta$  at phase $\theta$. Thus, by varying $\theta$ with a phase modulator, one can measure various components of quadrature phase amplitudes. Quantum state tomography is then performed using data from these measurements, employing methods like maximum likelihood estimation, to obtain the density matrix of the generated state \cite{Lvovsky2009}.

\subsubsection*{Experimental apparatus}
\begin{figure}[ht]
    \centering
    \includegraphics[width=1\linewidth]{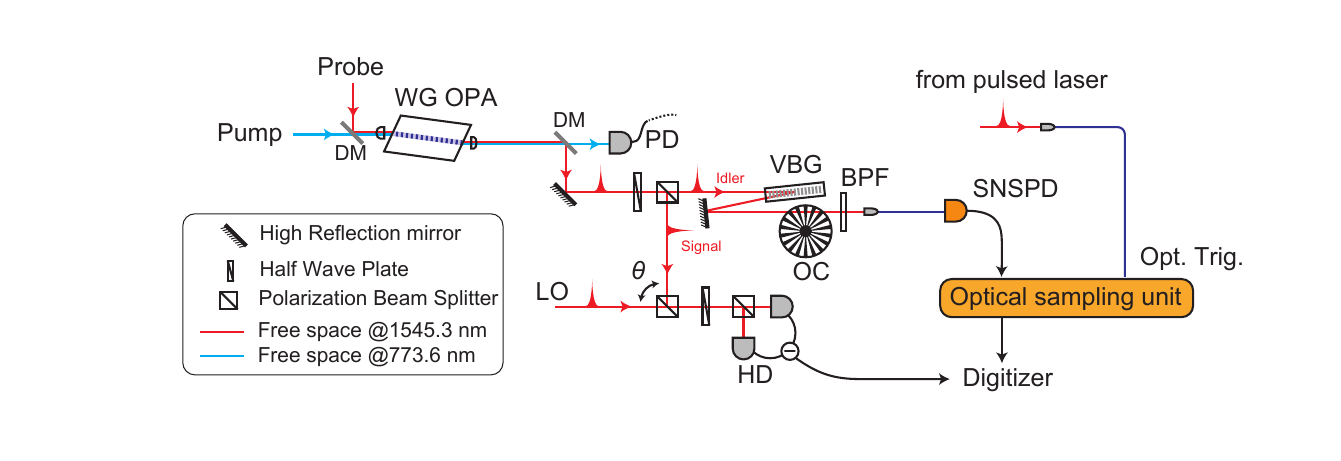}
    \caption{The experimental apparatus of photon subtraction. DM, dichroic mirror; WG OPA, waveguide optical parametric amplifier; PD, photodiode; VBG, volume Bragg grating; OC, optical chopper; BPF, optical band-pass filter; SNSPD, superconducting nanostrip photon detector; LO, local oscillator light; HD, homodyne detector. \black{One of the pulsed laser outputs is used for the optical trigger (Opt. Trig.). }}
    \label{fig:photon_subt}
\end{figure}

The experimental apparatus is depicted in Fig.~\ref{fig:photon_subt}. A 10-mm long periodically-polled lithium niobate waveguide (waveguide OPA: WG OPA), pumped by a pulsed laser with a central wavelength of 772.66 nm and a pulse duration of 10 ps, generates pulsed squeezed light. The pump light and squeezed light are separated by a dichroic mirror (DM), and the squeezed light is split into two paths (Signal and Idler) by a variable beam splitter consisting of a half-wave plate (HWP) and a Polarizing Beam Splitter (PBS). The transmitted light (Idler) through the PBS passes through a volume Bragg grating (VBG) with a bandwidth of 30 GHz before being injected into a single-mode fiber and sent to the SNSPD, where it is detected using the previously mentioned optical sampling method.

The quadrature-phase amplitude of the reflected light from the PBS (Signal) is measured with a homodyne detector (HD) using pulsed light as the LO. To prevent light reflected from surfaces such as the photodiodes' surface and lenses used in the HD from entering the SNSPD, we appropriately insert a Faraday rotator and pinhole (not shown). Still, some return light that cannot be removed is detected by the SNSPD. In our experimental system, the target signal and unnecessary detection signals are separated by a few nanoseconds, thus ignored by the optical sampling method thanks to its high temporal resolution. The signals from the HD and SNSPD are acquired by a high-speed digitizer (Keysight, M5200A). The built-in field-programmable gate array in the digitizer classifies events where the optical sampling method detects the desired photons and records the corresponding homodyne measurement results.

In the homodyne measurement, we lock the phase with the LO $\theta$, so we introduce probe light for phase reference. To prevent the probe light from entering the SNSPD, we employ a sample-and-hold method \cite{Endo2023}. The phase of the squeezed light generated by the WG OPA is determined by the pump light. When the probe light is injected into the WG OPA, the intensity of each light changes according to the phase relationship between the probe and pump lights, and that phase relationship can be locked based on these signals. In our case, the pump light transmitted through the OPA was received by a photodiode (PD) and used for locking. In this method, there is no optical element for pickup (typically ~1\%) in the optical path of the squeezed light, so the loss is reduced. The long-term and short-term phase fluctuations are negligible during the experiment.

In this experiment, the phase of the homodyne measurement was set to eight bases ($-67.5$, $-45.0$, $-22.5$, $0$, $22.5$, $45.0$, $67.5$, and $90.0$ degrees). Here, 0 degrees is defined as the anti-squeeze and 90 degrees as the squeeze. The signal with the pump light blocked was also measured and used as the shot noise data. The quadrature phase amplitudes were measured in each base, and the density matrix was obtained by the maximum likelihood estimation. Uncertainty was evaluated by the bootstrap method. Note that, throughout this paper, no corrections/post-
processing for measurement imperfections (e.g. optical losses, (spatial, temporal, and frequency) mode mismatches, electrical noises, detector efficiencies, dark counts of the SNSPD, and so on) has been made.

\subsubsection*{Quantum state tomography of generated states}
We generated and evaluated three types of quantum states for the following purposes: (1) to demonstrate the utility of the optically-sampled SNSPD as a high-performance single-photon detector; (2) to demonstrate the enhancement of non-classicality through the use of the SNSPD-based PNRD; (3) to provide an example of states that could not be generated without PNRDs.

\begin{figure}[ht]
    \centering
    \includegraphics[width=1\linewidth]{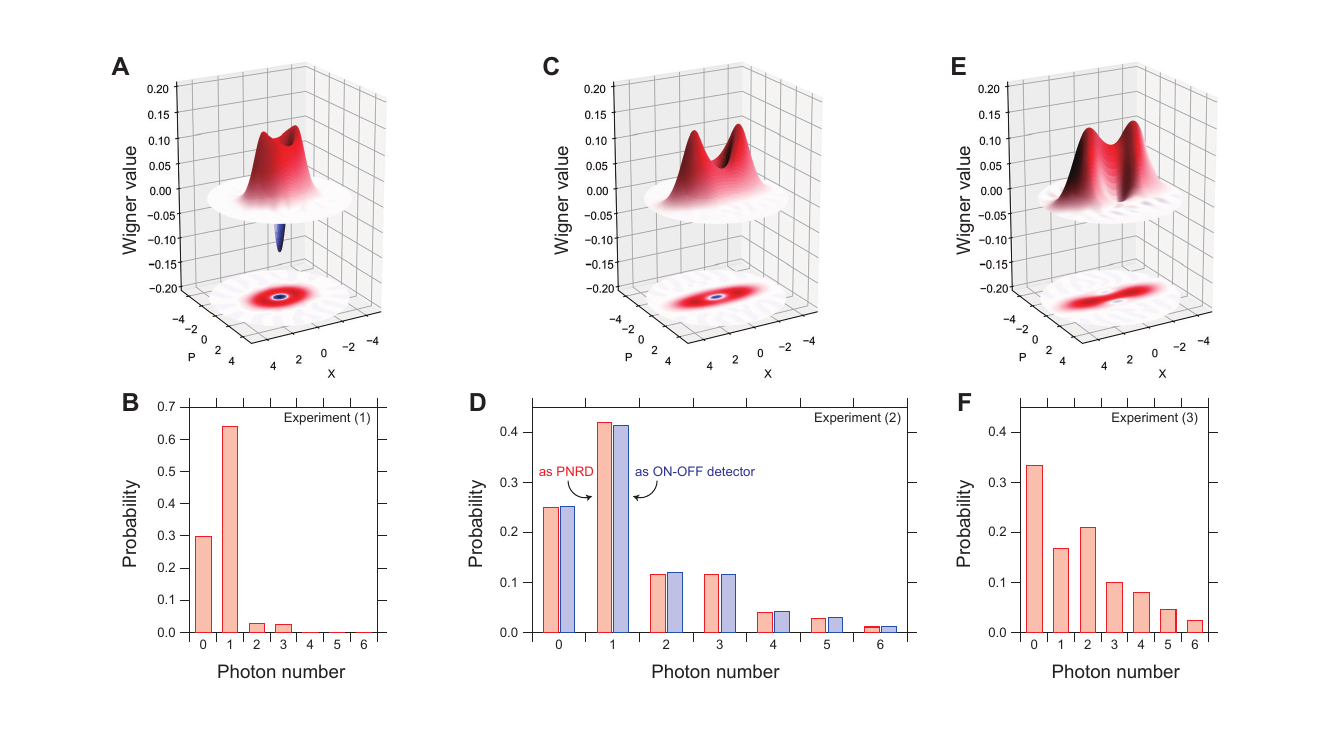}
    \caption{Experimental results. The top row (A, C, and E) shows Wigner functions, and the bottom row (C, D, and F) shows photon number distributions. A and B, single-photon subtraction (experiment (1)). C and D, Comparison of the SNSPD as a ON-OFF detector vs. PNRD in single-photon subtraction (experiment (2)). E and F, two-photon subtraction (experiment (3)). The error bars are small enough that they are not included in the plots. See Tab.~\ref{tab:photonnumberdist} for detailed numbers.}
    \label{fig:result_cat}
\end{figure}

\paragraph*{Experiment (1)}
In the first experiment, we set the peak power of the pump pulses entering the WG OPA to approximately 100 mW and the tap by the variable beam splitter to 2\%. At this setting, the squeezed light after the tap, measured by the HD, exhibited a squeezing level of 1.27 dB and an anti-squeezing level of 1.49 dB. The signal from the SNSPD was acquired using optical sampling, and about 100,000 quadrature phase amplitude data points were measured at each measurement phase, totaling around 800,000 points. The Wigner function and photon number distribution plotted from the obtained density matrix are shown in Fig.~\ref{fig:result_cat} A and B. Firstly, the Wigner function's value at the origin was $-0.108(2)$, indicating strong non-classicality. Figure~\ref{fig:result_cat} B shows a prominent single-photon component ($0.642(3)$), clearly resulting from the detection of one photon from a squeezed light comprising even-numbered photons. Note that the photon number distributions for all experiments are shown in Tab.~\ref{tab:photonnumberdist}. In this experiment, the count rate was around 400 count-per-second (cps). 
Despite the influence of factors like dark counts contributing to over 100 cps in this experiment, the use of the optical sampling method has enabled the selective detection of only the events of interest, effectively reducing the stray count to less than 1 cps.

Homodyne measurement of states generated by ultra-short pulses is generally challenging, tending to degrade the negative values in the Wigner function compared to continuous wave light experiments. This is due to the multi-mode nature of the quantum states in pulsed light and the need to match the time waveform of the state with the LO waveform during verification \cite{Slusher1987,Gerrits2010,Endo2023}. In our experiment, we minimized these effects using the low-loss WG OPA, appropriate wavelength filters, and LO waveform optimization, achieving an experimental setup comparable to a continuous wave system.

\paragraph*{Experiment (2)}
In the next experiment, we increased the peak power of the pump to approximately 1 W and set the tap ratio by the variable beam splitter to 16\%. The squeezing level was 2.93 dB, and the anti-squeezing level was 4.46 dB, respectively. With such high squeezing levels and larger tap ratios, the influence of the idler's two-photon or more photon components becomes non-negligible. In other words, if the SNSPD lacks photon number-resolving capability (like ON-OFF detectors), it would be impossible to distinguish between one and two photons. In contrast, with a PNRD, these can be distinguishable, and only single-photon events can be extracted. We collected all signals obtained by optical sampling, labeled the one and two-photon events with the digitizer, saved the quadrature phase amplitude values, and conducted the analysis. Approximately 800,000 quadrature amplitude data points were acquired at each phase, totaling around 6.4 million points, to calculate the density matrix.  The results are shown in Fig.~\ref{fig:result_cat} C and D. Fig.~\ref{fig:result_cat} C represents the Wigner function when the SNSPD is used as an ON-OFF detector (only one is plotted as it is difficult to discern the difference between using it as an ON-OFF detector and a PNRD in the Wigner function plot). Even as an ON-OFF detector, negative values near the origin were observed, clearly indicating a non-classical quantum state. However, the values at the origin when used as an ON-OFF detector and a PNRD were $-0.0443(5)$ and $-0.0486(6)$, respectively, showing improvement with the PNRD. Detecting two photons simultaneously increases the even component of the generated state, bringing the Wigner function's origin closer to positive. Figure~\ref{fig:result_cat} D compares the photon number distributions of both cases. Even in $n=1$ experiments, the use of PNRDs is significant. Conventionally, experiments could only be conducted in the weak pump limit, where components larger than one photon were negligible, by minimizing pump power and taps. PNRDs allow for high count rate state generation, moving beyond this limitation. 

\paragraph*{Experiment (3)}
Finally, from the same data as the previous experiment (2), we selected the two-photon detection events and performed quantum tomography. As evident from the Wigner function in Fig.~\ref{fig:result_cat} E, the origin is positive. Furthermore, $W(0,-0.92)=-0.005(2)$ and $W(0,0.90)=0.011(2)$, showing negative values and indicating strong non-classicality. As seen in the photon number distribution in Fig.~\ref{fig:result_cat} F, the even components are significantly larger. These results demonstrate the effectiveness of SNSPD-based PNRDs in detecting photons larger than one.

\begin{table}[ht]
\caption{Photon number distributions of generated quantum states in the experiments (1)-(3)}
\label{tab:photonnumberdist}
\begin{tabular}{cllll}
\hline
 & \multicolumn{4}{c}{Generated quantum states} \\ \cline{2-5} 
Photon number & (1) & (2) as ON-OFF & (2) as PNRD & (3) \\ \hline
0 & $2.99(2)\times10^{-1}$ & $2.523(3)\times10^{-1}$ & $2.499(3)\times10^{-1}$ & $3.34(2)\times10^{-1}$ \\
1 & $6.42(3)\times10^{-1}$ & $4.138(4)\times10^{-1}$ & $4.211(4)\times10^{-1}$ & $1.69(3)\times10^{-1}$ \\
2 & $2.82(29)\times10^{-2}$ & $1.201(5)\times10^{-1}$ & $1.174(5)\times10^{-1}$ & $2.10(3)\times10^{-1}$ \\
3 & $2.69(15)\times10^{-2}$ & $1.156(4)\times10^{-1}$ & $1.161(4)\times10^{-1}$ & $1.00(2)\times10^{-1}$ \\
4 & $2.52(52)\times10^{-3}$ & $4.179(28)\times10^{-2}$ & $4.071(36)\times10^{-2}$ & $8.04(20)\times10^{-2}$ \\
5 & $9.54(194)\times10^{-4}$ & $2.989(25)\times10^{-2}$ & $2.937(27)\times10^{-2}$ & $4.58(16)\times10^{-2}$ \\
6 & $2.11(89)\times10^{-4}$ & $1.194(19)\times10^{-2}$ & $1.161(21)\times10^{-2}$ & $2.54(11)\times10^{-2}$ \\ \hline
\end{tabular}
\end{table}

\section*{Discussions}
Until now, SNSPDs have been recognized as promising PNRDs due to their high speed and high detection efficiency. However, their speed necessitated high-temporal-resolution measurement devices, leaving them just a step short of practical application. Particularly in the field of continuous-variable optical quantum information processing, there has been a strong demand for fast PNRDs like SNSPDs for high generation rates of non-classical states like GKP logical qubits.
In this paper, we concur this challenge with the optical sampling method, facilitating easier real-time photon number resolving. Furthermore, this technique is directly applicable to jitter analysis in various photodetectors, including SNSPDs. For instance, the temporal resolution of state-of-the-art SNSPDs reaches sub-3 ps \cite{Korzh2020}, approaching the limits of electrical measurement equipment. By optimizing the pulse duration and dispersion of the pulse laser, our developed method's temporal resolution can potentially reach sub-picosecond levels, opening a wide range of applications from elucidating photon detection mechanisms in SNSPDs to LIDAR applications with ultra-weak light.

In this paper, we specifically applied the method to generating non-classical quantum states, a field where real-time, true photon number resolution is crucial. We generated approximate states of Schr\"{o}dinger cat by subtracting photons from squeezed light emitted from the OPA. The high temporal resolution of the SNSPD-type PNRD led to observed improvements in the non-classicality of the states, specifically in the negative values of the Wigner function. Additionally, we succeeded in generating quantum states that could not be produced without a PNRD. Although the SNSPD used in this experiment could not detect more than three photons, using an appropriate SNSPD would enable the detection of a greater number of photons. Moreover, an array of PNRDs, rather than a single-photon detector array, has significant implications \cite{Harder2016,Eaton2022}. An array of spatially multiplexed PNRDs could potentially increase the maximum measurable photon number and our method can be easily extended to spatially-multiplexed SNSPD.

Furthermore, in this paper, the generation rate was limited by the laser's repetition frequency of 5 MHz. Since the temporal resolution of the SNSPD used is approximately 60 ps, in principle, using a pulse laser with a 5 GHz repetition frequency could increase the generation rate by a thousandfold.

Thus, this method elevates SNSPD-type PNRDs from a primarily experimental verification level to a practically applicable level, significantly expanding the range of applications for PNRDs.

\section*{Materials and Methods}
\subsection*{Details of the SNSPD}
The SNSPD used in our experiment employs niobium titanium nitride (NbTiN) as the superconducting material. The device's active area is $25\times25\ \mathrm{\mu m}^2$, featuring a nanostrip with a width of 80 nm and a meander structure with a 100 nm gap. Enhanced light absorption efficiency is achieved through a cavity structure \cite{Miki2017}, and light is directed to the active area using a single-mode fiber (SMF-28). The bias current is applied through a low-noise voltage source (Stanford Research Systems, SIM928) and a bias tee in the cryogenic amplifier (see \cite{Endo2021}). At a bias current of 25 microamperes, the detection efficiency of the device was approximately 60\%, with a dark count rate of about 100 cps.

\subsection*{Light sources and phase locking for nonclassical state generation}
In this experiment, we utilized a femtosecond mode-locked laser with a repetition frequency of 100 MHz (MenloSystems, ULN-Comb) as the source for generating pump light, local oscillator (LO) light, probe light, and the optical trigger. The basic setup is largely similar to our previous publication \cite{Endo2023}.

The original laser features a repetition frequency of 100 MHz, a central wavelength of 1550 nm, and a pulse duration of about 100 fs. One of its multiple output ports is used to generate pump and probe lights. By employing an AOM with a driving frequency of 400 MHz for pulse picking, followed by dual-stage EDFAs, which are home-made and commercial (PriTel, HPP-PMFA-22) ones, and spectral filters, we produce pulses with a repetition frequency of 5 MHz, a pulse width of 6 ps, and a central wavelength of 1545.32 nm (194.0 THz). These pulses are directed into a bulk PPMgSLT crystal (Oxide, fan-out PPMgSLT) for second-harmonic generation. A DM separates the fundamental and second-harmonic lights. The fundamental light undergoes chopping at a frequency of 500 Hz using two AOMs. A piezoelectric actuator is attached to one of the mirrors in the probe light path, used for locking the probe light against the squeezed light.

The second harmonic's spectral width is limited to 30 GHz using a 4f optical system spectrometer and slit as a frequency filter. Bandpass and shortpass filters are placed to prevent the fundamental light components from being separated by the DM and the leaked EDFA's pump light at the wavelength of 976 nm. The light then enters a polarization-maintaining fiber after passing through an optical delay line used for timing adjustment.

For the LO light, we use another output port of the laser. After limiting the spectral width to 200 GHz with a bandpass filter, the light undergoes pulse picking via an AOM and amplification with dual-stage EDFAs (Thorlabs, EDFA100P). To control the phase for the homodyne measurement, we use a waveguide-type phase modulator and a home-made fiber stretcher to compensate long-term drift. A commercial waveshaper shapes the LO light's temporal waveform, limits its spectral width to 30 GHz, and compensates for dispersion. An optical fiber delay is incorporated to synchronize the generated state with the LO light.

Power fluctuations in the probe, pump, and LO light can lead to measurement errors, so we have implemented power stabilization mechanisms for each.

The phase of the squeezed light generated in the WG-OPA is determined by the phase of the pump light. The phase relationship between the pump and probe lights appears as modulation in the pump (or probe) light transmitted through the WG-OPA, which we detect using the PD and feedback control to the PZT. The phase between the probe and LO lights can be detected by the HD, with the signal fed back to the waveguide phase modulator inserted in the LO light path. As mentioned, the probe light is chopped at 500 Hz and uses a sample-and-hold method. The phase fluctuation during the hold is kept below one degree, not affecting this experiment's outcome.

\subsection*{Loss budget in the quantum state generation and verification}
The measured losses of the quantum state generation experiment can be found in Tab.~\ref{tab:loss}. We stress that we did not correct these losses in the tomography results. 
\begin{center}
\begin{table}[ht]
\caption{Loss budget}
\label{tab:loss}
\begin{tabular}{ll}
\hline
Element                & Loss (\%) \\ \hline
WG OPA                 & 5         \\
Inefficiency of HD     & 4         \\
Spatial mode mismatch  & 8        \\
Temporal mode mismatch & 4         \\
Propagation loss       & 5         \\
Circuit noise          & 3.5       \\
Inefficiency of SNSPD  & 40        \\
Dark and fake count contributions of SNSPD & $<$1 \\
Fiber coupling loss for Idler    & 15        \\ \hline
\end{tabular}
\end{table}
\end{center}


\bibliography{scibib2}

\bibliographystyle{Science}

\section*{Acknowledgments}
\paragraph*{Funding}
This work was partly supported by Japan Science and Technology (JST) Agency (Moonshot R \& D) Grant No. JPMJMS2064 and JPMJMS2066, UTokyo Foundation, and donations from Nichia Corporation of Japan. M. E. acknowledges the funding from JST PRESTO (JPMJPR2254). W.A. acknowledges the funding from Japan Society for the Promotion of Science (JSPS) KAKENHI (No. 23K13040). K.T. acknowledges the funding from JSPS KAKENHI (No. 23K13038, 22K20351). \black{T.S. acknowledges the funding from  JSPS KAKENHI (No. 23KJ0518).} W.A., K.T, and M.E. acknowledge supports from Research Foundation for OptoScience and Technology. 

\paragraph*{Author contributions}
M.E. conceived and led the experiment. M.E., K. Takahashi and T.S. installed and evaluated the SNSPD-type PNRD with optical sampling. M.E. and T.N. constructed the experimental system and coded the measurement and analysis programs. T.K., A.I., and T.U. provided the OPA used in the experiment. F.C., M.Y., S.M., and H.T. provided the SNSPD used in this experiment. M.E., K. Takahashi, T.S., and T.N. analyzed the experimental results with the supervision of R.N., K. Takase, W.A., and A.F.  M.E. wrote the manuscript with the assistance of all the other authors.

\end{document}